\documentclass[12pt]{article}
\usepackage{graphicx}

\begin{document}

\baselineskip=18pt plus 1pt minus 1pt
\begin{center} 

{\Large\bf Critical point symmetries in nuclei}

\bigskip\bigskip 
Dennis Bonatsos $^a$, D. Lenis $^a$, D. Petrellis $^a$, P. A. Terziev $^b$,
I. Yigitoglu $^c$

\bigskip 
$^a$ Institute of Nuclear Physics, N.C.S.R. ``Demokritos'',\\
 GR-15310 Aghia Paraskevi, Attiki, Greece

$^b$ Institute for Nuclear Research and Nuclear Energy, Bulgarian
Academy of Sciences,\\
 72 Tzarigrad Road, BG-1784 Sofia, Bulgaria

$^c$ Hasan Ali Yucel Faculty of Education, Istanbul University,\\
TR-34470 Beyazit, Istanbul, Turkey

\bigskip\bigskip 
{\bf Abstract}

\end{center}
\medskip 
Critical Point Symmetries (CPS) appear in regions of the nuclear chart 
where a rapid change from one symmetry to another is observed. The first 
CPSs, introduced by F. Iachello, were E(5), which corresponds to the 
transition  from vibrational [U(5)] to $\gamma$-unstable [O(6)] behaviour,
and X(5), which represents the change from vibrational [U(5)] to prolate 
axially deformed [SU(3)] shapes. These CPSs have been obtained as special 
solutions of the Bohr collective Hamiltonian. More recent special solutions
of the same Hamiltonian, to be described here, include Z(5) and Z(4), 
which correspond to maximally triaxial shapes (the latter with ``frozen''
$\gamma=30^{\rm o}$), as well as X(3), which corresponds to prolate shapes 
with ``frozen'' $\gamma=0^{\rm o}$. CPSs have the advantage of providing 
predictions which are parameter free (up to overall scale factors) and 
compare well to experiment. However, their mathematical structure 
[with the exception of E(5)] needs to be clarified. 

\newpage 

\section{Introduction} 

Critical point symmetries \cite{IacE5,IacX5}, describing nuclei at 
points of shape phase transitions between different limiting symmetries,
have recently attracted considerable attention, since they lead to 
parameter independent (up to overall scale factors) predictions which 
are found to be in good agreement with experiment 
\cite{CZE5,ClarkE5,Zamfir,CZX5,ClarkX5}. The E(5) 
critical point symmetry \cite{IacE5} is supposed 
to correspond to the transition from vibrational [U(5)] to $\gamma$-unstable 
[O(6)] nuclei, while the X(5) critical point symmetry \cite{IacX5}
is assumed to describe 
the transition from vibrational [U(5)] to prolate axially 
symmetric [SU(3)] nuclei. Both symmetries are obtained as special solutions 
of the Bohr Hamiltonian \cite{Bohr}. In the E(5) case \cite{IacE5} 
the potential is supposed to 
depend only on the collective variable $\beta$ and not on $\gamma$. 
Then exact separation of variables is achieved and the equation containing 
$\beta$ can be solved exactly \cite{IacE5,Wilets} for an infinite square well 
potential in $\beta$, the eigenfunctions being Bessel 
functions of the first kind, while the equation containing the angles has been 
solved a long time ago by B\`es \cite{Bes}. 
In the X(5) case \cite{IacX5}
the potential is supposed to be of the form $u(\beta)+u(\gamma)$.
Then approximate separation of variables is achieved in the special case
of $\gamma \simeq 0$, the $\beta$-equation with an infinite square 
well potential 
leading to Bessel eigenfunctions, while the $\gamma$-equation with a harmonic 
oscillator potential having a minimum at $\gamma=0$ leads to a two-dimensional
harmonic oscillator with Laguerre eigenfunctions \cite{IacX5}. 
In both cases the full five variables 
of the Bohr Hamiltonian \cite{Bohr} (the collective variables $\beta$ and 
$\gamma$, 
as well as the three Euler angles) are involved. The algebraic structure 
of E(5) is clear, since the Hamiltonian is the second order Casimir operator 
of E(5), which corresponds to the square of the momentum operator in five 
dimensions (see \cite{E5,Z4} and references therein), 
while an SO(5) subalgebra (generated by the angular momentum operators 
in five dimensions) exists. The algebraic structure of X(5) (if any, since 
X(5) is an approximate and not an exact solution) has not been
identified yet.  

It is of interest to identify additional special cases leading to analytical 
solutions of the Bohr Hamiltonian, and to examine their relation to 
critical behaviour of nuclei, clarifying in parallel their algebraic 
structure. 

It has been known for a long time that the Bohr equation gets simplified 
in the special case of $\gamma=30^{\rm o}$ \cite{Brink,MtVNPA}, since two of 
the principal moments
of inertia become equal in this case, guaranteeing the existence of a good 
quantum number (the projection $\alpha$ of angular momentum on the body-fixed 
$\hat x'$ axis), although the nucleus possesses a triaxial shape. In other 
words, the Hamiltonian possesses a symmetry, while the shape of the nucleus 
does not.  By allowing
the potential to be of the form $u(\beta)+u(\gamma)$, and by permitting 
$\gamma$ to vary only around $\gamma\simeq 30^{\rm o}$, approximate 
separation of variables is achieved \cite{Z5}, similar in spirit to the X(5) 
solution. The $\beta$-equation with an infinite square well potential leads 
then to Bessel eigenfunctions, while the $\gamma$-equation with a harmonic 
oscillator potential having a minimum at $\gamma=30^{\rm o}$ takes the 
form of a simple harmonic oscillator equation. The full five variables of the 
Bohr Hamiltonian are involved in this case, while the algebraic structure 
(if any, since the solution is approximate) is yet unknown. This solution,
which has been called Z(5) \cite{Z5}, is presented in Section 2. 

Separation of variables becomes exact by ``freezing'' the $\gamma$ variable 
to the special value of $\gamma=30^{\rm o}$, in the spirit of the 
Davydov and Chaban \cite{Chaban} approach. Then the $\beta$-equation with 
an infinite square well potential leads to Bessel eigenfunctions \cite{Z4}, 
while 
the equation involving the Euler angles and the parameter $\gamma$ (which is
not a variable any more) leads to the solution obtained by Meyer-ter-Vehn 
\cite{MtVNPA}. The  projection $\alpha$ of angular momentum on the body-fixed 
$\hat x'$ axis is a good quantum 
number also in this case. Only four 
variables ($\beta$ and the three Euler angles) are involved,
while the full algebraic structure is yet unknown. It has been remarked
\cite{Z4}, however, that the ground state band of this model coincides with 
the ground state band of E(4), the Euclidean algebra in four dimensions. 
This solution, which has been labelled as Z(4) \cite{Z4}, is presented 
in Section 3. 

The question arises then of what happens in the case one ``freezes'' the 
$\gamma$ variable to the value $\gamma=0$, which corresponds to axially 
symmetric prolate shapes, for which the projection $K$ of angular momentum 
on the body-fixed $\hat z$-axis is a good quantum number. It turns out 
\cite{X3} that 
only three degrees of freedom are relevant in this case, since the nucleus 
is axially symmetric, so that two angles suffice for describing its direction 
in space, while the variable $\beta$ describes its shape. Separation 
of variables becomes exact \cite{X3}, the $\beta$ equation with an infinite 
square 
well potential leading to Bessel eigenfunctions, while the equation involving 
the angles leads to the simple spherical harmonics. The algebraic structure 
of this model is yet unknown. This solution, which  has been called X(3) 
\cite{X3}, is presented in Section 4. 
  
Finally, in Section 5 the present results are briefly discussed and plans 
for further work are exposed. 

\section{The Z(5) model}  

The original Bohr Hamiltonian \cite{Bohr} is
$$ H = -{\hbar^2 \over 2B} \left[ {1\over \beta^4} {\partial \over \partial 
\beta} \beta^4 {\partial \over \partial \beta} + {1\over \beta^2 \sin 
3\gamma} {\partial \over \partial \gamma} \sin 3 \gamma {\partial \over 
\partial \gamma} \right.$$
\begin{equation}\label{eq:e1}
\left. - {1\over 4 \beta^2} \sum_{k=1,2,3} {Q_k^2 \over \sin^2 
\left(\gamma - {2\over 3} \pi k\right) } \right]  +V(\beta,\gamma),
\end{equation}
where $\beta$ and $\gamma$ are the usual collective coordinates, while
$Q_k$ ($k=1$, 2, 3) are the components of angular momentum in the intrinsic 
coordinate system and $B$ is the mass parameter.  

In the case in which the potential 
has a minimum around $\gamma =\pi/6$ one can write  the last term of Eq. 
(\ref{eq:e1}) in the form  $ 4(Q_1^2+Q_2^2+Q_3^2)-3Q_1^2$. 
Using this result in the Schr\"odinger equation corresponding to 
the Hamiltonian of Eq. (\ref{eq:e1}), introducing \cite{IacE5} reduced energies
 $\epsilon = 2B E /\hbar^2$ and reduced potentials $u = 2B V /\hbar^2$,  
and assuming \cite{IacX5} that the reduced potential can be separated into two 
terms, one depending on $\beta$ and the other depending on $\gamma$, i.e. 
$u(\beta, \gamma) = u(\beta) + u(\gamma)$, the Schr\"odinger equation can 
be separated into two equations 
\begin{equation} \label{eq:e3}
\left[ -{1\over \beta^4} {\partial \over \partial \beta} \beta^4 
{\partial \over \partial \beta} + {1\over 4 \beta^2} (4L(L+1)-3\alpha^2)  
+u(\beta) \right] \xi_{L,\alpha}(\beta) =\epsilon_\beta  \xi_{L,\alpha}(\beta),
\end{equation}
\begin{equation}\label{eq:e4} 
\left[ -{1\over \langle \beta^2\rangle \sin 3\gamma} {\partial \over 
\partial \gamma}\sin 3\gamma {\partial \over \partial \gamma} 
+u(\gamma)\right] \eta(\gamma) = 
\epsilon_\gamma \eta(\gamma),
\end{equation}
where $L$ is the angular momentum quantum number, $\alpha$ is the projection 
of the angular momentum on the body-fixed $\hat x'$-axis 
($\alpha$ has to be an even integer \cite{MtVNPA}),
$\langle \beta^2 \rangle$ is the average of $\beta^2$ over $\xi(\beta)$, 
and $\epsilon= \epsilon_\beta +\epsilon_\gamma$. 

The total wave function should have the form 
$ \Psi(\beta,\gamma,\theta_i) = \xi_{L,\alpha}(\beta) \eta(\gamma) 
{\cal D}^L _{M,\alpha}(\theta_i)$, 
where $\theta_i$ ($i=1$, 2, 3) are the Euler angles, ${\cal D}(\theta_i)$ 
denote Wigner functions of them, $L$ is the angular momentum quantum number, 
while $M$ and $\alpha$ are the eigenvalues of the projections 
of angular momentum on the laboratory fixed $\hat z$-axis and the body-fixed 
$\hat x'$-axis respectively. 

Instead of the projection $\alpha$ of the angular momentum on the 
$\hat x'$-axis, it is customary to introduce the wobbling quantum number 
\cite{MtVNPA,BM} $n_w=L-\alpha$. Inserting $\alpha=L-n_w$ in 
Eq. (\ref{eq:e3}) one obtains 
$$ \left[ -{1\over \beta^4} {\partial \over \partial \beta} \beta^4 
{\partial \over \partial \beta} + {1\over 4 \beta^2} (L(L+4)+3n_w(2L-n_w))  
+u(\beta) \right] \xi_{L,n_w}(\beta)  $$
\begin{equation}\label{eq:e6}
 =\epsilon_\beta  \xi_{L,n_w}(\beta), 
\end{equation}
where the wobbling quantum number $n_w$ labels a series of bands 
with  $L=n_w,n_w+2,n_w+4, \dots$ (with $n_w > 0$) next to the ground state 
band (with $n_w=0$) \cite{MtVNPA}.  

In the case in which $u(\beta)$ is an infinite well potential
\begin{equation}\label{eq:e7} 
 u(\beta) = \left\{ \begin{array}{ll} 0 & \mbox{if $\beta \leq \beta_W$} \\
\infty  & \mbox{for $\beta > \beta_W$} \end{array} \right. ,  
\end{equation} 
one can use the transformation \cite{IacX5} 
$\tilde \xi(\beta) = \beta^{3/2} \xi(\beta)$, as well as the definitions 
\cite{IacX5} $\epsilon_\beta= k_\beta^2$, $z=\beta k_\beta$, in order 
to bring Eq. (\ref{eq:e6}) into the form of a Bessel equation 
\begin{equation}\label{eq:e8}
{d^2 \tilde \xi \over d z^2} + {1\over z} {d \tilde \xi \over d z} 
+ \left[ 1 - {\nu^2 \over z^2}\right] \tilde \xi=0,
\end{equation}
with 
\begin{equation}\label{eq:e9} 
\nu = {\sqrt{4L(L+1)-3\alpha^2+9}\over 2}= 
{\sqrt{L(L+4)+3n_w(2L-n_w)+9}\over 2}.   
\end{equation}

\begin{figure}[htb]
\includegraphics[height=100mm]{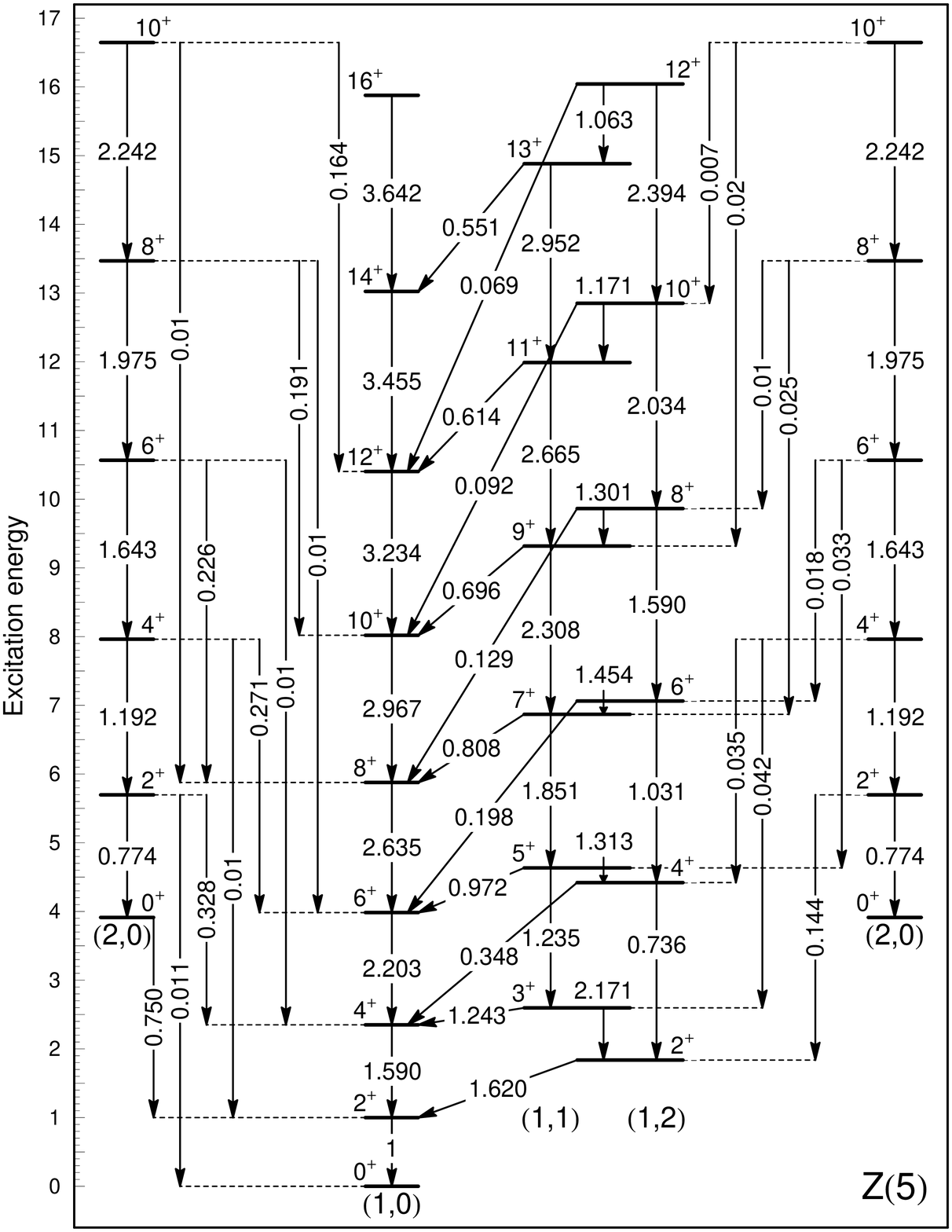} 
\includegraphics[height=100mm]{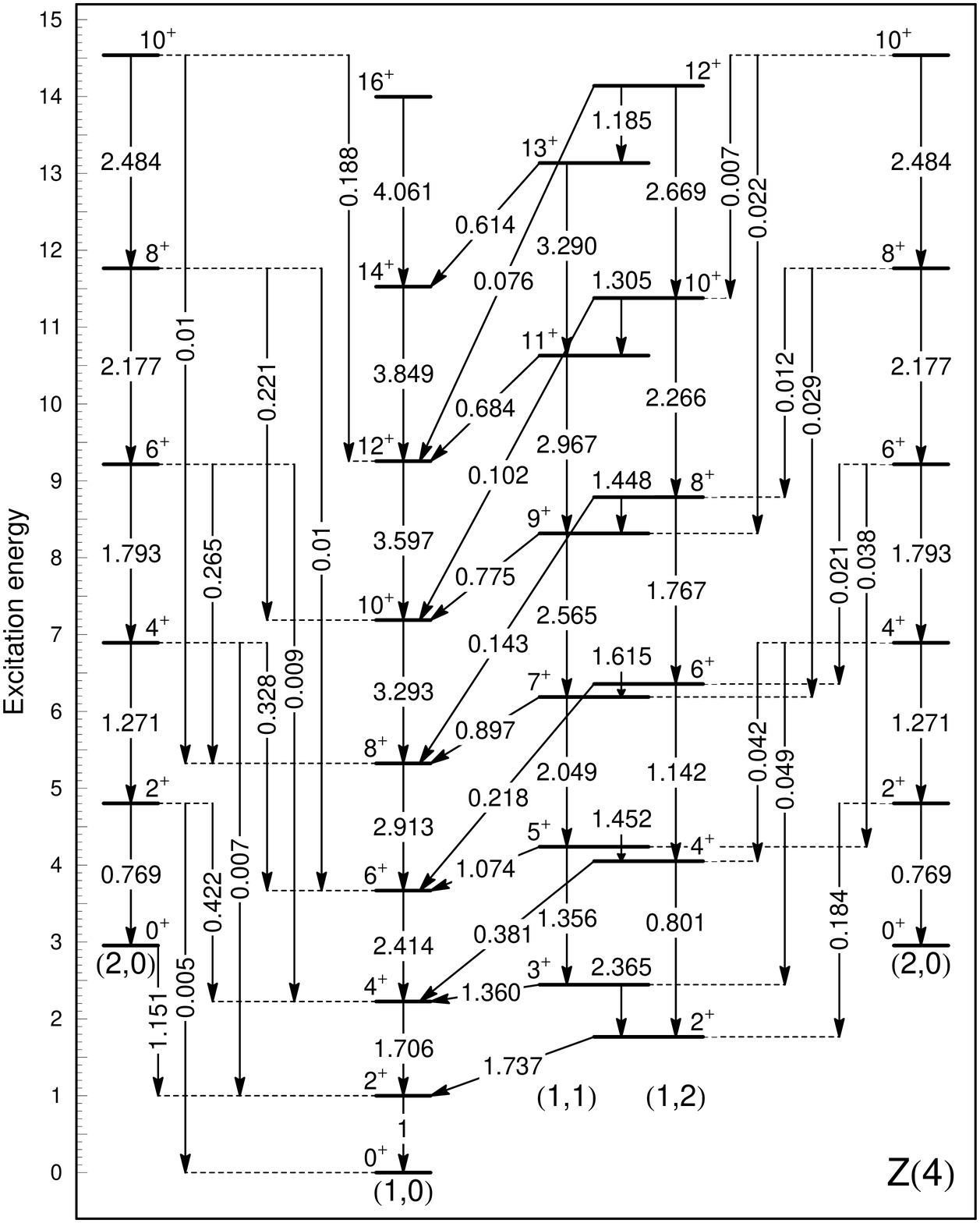} 
\caption{(a)
Intraband and interband B(E2) transition rates in the Z(5) 
model \cite{Z5}, normalized to the B(E2;$2_{1,0}\to 0_{1,0}$) rate. Bands are 
labelled by $(s,n_w)$, their levels being normalized to $2_{1,0}$.  
The (2,0) band is shown both at the left and at 
the right end of the figure for drawing purposes.  
(b) Same for the Z(4) model \cite{Z4}. 
}\label{fig1}
\end{figure}

Then the boundary condition $\tilde \xi(\beta_W) =0$ 
determines the spectrum 
\begin{equation}\label{eq:e10}
\epsilon_{\beta; s,\nu} = \epsilon_{\beta; s,n_w,L} 
= (k_{s,\nu})^2, \qquad k_{s,\nu} = x_{s,\nu}/ \beta_W, 
\end{equation}
and the eigenfunctions 
\begin{equation}\label{eq:e11} 
\xi_{s,\nu}(\beta) = \xi_{s,n_w,L} (\beta)= \xi_{s,\alpha,L}(\beta)= 
c_{s,\nu} \beta^{-3/2} J_\nu (k_{s,\nu} \beta), 
\end{equation}
where $x_{s,\nu}$ is the $s$th zero of the Bessel function $J_\nu(z)$, 
while the normalization constants $c_{s,\nu}$ are determined from the 
normalization condition $ \int_0^\infty \beta^4 \xi^2_{s,\nu}(\beta) 
d\beta=1$. The notation for the roots has been kept the same as in Ref. 
\cite{IacX5}, while for the energies the notation $E_{s,n_w,L}$ 
will be used. The ground state band corresponds to $s=1$, $n_w=0$.
We shall refer to the model corresponding to this solution as Z(5)
(which is not meant as a group label), in analogy to the E(5) \cite{IacE5} 
and X(5) \cite{IacX5} models.  

The $\gamma$-part of the spectrum is obtained from Eq. (\ref{eq:e4}), 
which, in the case of a harmonic oscillator potential having a 
minimum at $\gamma =\pi/6$, takes the form of a simple harmonic 
oscillator equation \cite{Z5}. Similar potentials and solutions in the 
$\gamma$-variable have been considered in \cite{Bohr}.
The total energy in the case of the Z(5) model is then
\begin{equation}\label{eq:e24}
E(s,n_w,L,n_{\tilde\gamma}) = E_0 + A (x_{s,\nu})^2 + B n_{\tilde \gamma}, 
\end{equation}
where $n_{\tilde \gamma}$ is the quantum number of the oscillator occuring 
in the $\gamma$-equation. 

The calculation of B(E2) transition rates has been described in detail 
in Ref. \cite{Z5}. The resulting level scheme is shown in Fig. 1(a).
Experimental manifestations of Z(5) seem to appear in the $^{192-196}$Pt 
region \cite{Z5}.   

\section{The Z(4) model}   

\subsection{The Z(4) solution}   

In the model of Davydov and Chaban \cite{Chaban} it is assumed that the 
nucleus is rigid with respect to $\gamma$-vibrations. Then the Hamiltonian 
depends on four variables ($\beta,\theta_i$) and has the form \cite{Chaban} 
\begin{equation}\label{eq:z41} 
H = -\frac{\hbar^2}{2B}\Biggl[\frac{1}{\beta^3}\frac{\partial}{\partial\beta}
\beta^3\frac{\partial}{\partial\beta} - \frac{1}{4\beta^2}
\sum_{k=1}^{3}\frac{Q_{k}^2}{\sin^2(\gamma-\frac{2\pi}{3}k)}\Biggr]
+ U(\beta).
\end{equation}
In this Hamiltonian $\gamma$ is treated as a parameter and 
not as a variable. The kinetic energy term of Eq. (\ref{eq:z41}) is different 
from the one appearing in the E(5) and X(5) models, because of the different 
number of degrees of freedom treated in each case (four in the former case, 
five in the latter). 
 
Introducing \cite{IacE5} reduced energies $\epsilon= (2 B/\hbar^2) E$ and 
reduced 
potentials $u=(2B/\hbar^2) U$, and considering a wave function of the form 
$\Psi(\beta,\theta_i)=\phi(\beta)\psi(\theta_i)$, where $\theta_i$ (
$i=1$, 2, 3) are the Euler angles, 
separation of variables 
leads to two equations
\begin{equation}\label{eq:z42} 
\Biggl[\frac{1}{\beta^3}\frac{\partial}{\partial\beta}
\beta^3\frac{\partial}{\partial\beta} - \frac{\lambda}{\beta^2}
+(\epsilon-u(\beta))\Biggr] \phi(\beta) = 0, 
\end{equation}
\begin{equation}\label{eq:z43} 
\Biggl[\frac{1}{4}\sum_{k=1}^{3}\frac{Q_{k}^2}{\sin^2(\gamma-\frac{2\pi}{3}k)}
-\lambda \Biggr]\psi(\theta_i) = 0.
\end{equation}

In the case of $\gamma=\pi/6$, the last equation takes the form 
\begin{equation}\label{eq:z44}
\Biggl[\frac{1}{4}(Q_1^2 + 4Q_2^2 + 4Q_3^2)-\lambda \Biggr]\psi(\theta_i) = 0.
\end{equation}
This equation has been solved by Meyer-ter-Vehn \cite{MtVNPA}, the 
eigenfunctions being  
\begin{equation}\label{eq:z45}
\psi(\theta_i)=\psi^L_{\mu,\alpha}(\theta_i) =
\sqrt{\frac{2L+1}{16\pi^2(1+\delta_{\alpha,0})}}\,
\Bigl[{\cal D}^{(L)}_{\mu,\alpha}(\theta_i) 
+ (-1)^L {\cal D}^{(L)}_{\mu,-\alpha}(\theta_i)\Bigr]
\end{equation}
with 
\begin{equation}\label{eq:z46}
\lambda = \lambda_{L,\alpha} = L(L+1) -\frac{3}{4}\,\alpha^2, 
\end{equation}
where $\mu$ and $\alpha$ are the eigenvalues of the projections 
of angular momentum on the laboratory fixed $\hat z$-axis and the body-fixed 
$\hat x'$-axis respectively. $\alpha$ has to be an even integer 
\cite{MtVNPA}. As in the previous section, the wobbling quantum number,
$n_w$ \cite{MtVNPA,BM}, is introduced at this point. 

The ``radial'' Eq. (\ref{eq:z42}) is exactly soluble in the case of an 
infinite square well potential [Eq. (\ref{eq:e7})]. 
Using the transformation $\phi(\beta)=\beta^{-1}f(\beta)$, Eq. (\ref{eq:z42})
becomes a Bessel equation 
\begin{equation}\label{eq:z47}
\Biggl[\frac{\partial^2}{\partial\beta^2} +
\frac{1}{\beta}\frac{\partial}{\partial\beta}
+ \Bigl(\epsilon - \frac{\nu^2}{\beta^2} \Bigr)\Biggr] f(\beta) = 0, 
\end{equation}
with
\begin{equation}\label{eq:z48}
\nu=\sqrt{\lambda+1}
={\sqrt{L(L+4)+3 n_w(2L-n_w)+4}\over 2}.
\end{equation}
Then the boundary condition $f(\beta_W)=0$ determines the spectrum,
which is given by Eq. (\ref{eq:e10}), while the eigenfunctions are  
\begin{equation}\label{eq:z49}
\phi(\beta)=\phi_{s,\nu}(\beta) =\phi_{s,n_w,L}(\beta) 
= \frac{1}{\sqrt{c}}\,\beta^{-1}
J_{\nu}(k_{s,\nu}\beta), \quad 
c = \frac{\beta_W^2}{2}\,J^2_{\nu+1}(x_{s,\nu})
\end{equation}
where the normalization constant $c$ is determined from the condition
$\int_0^{\beta_W} \beta^3 \phi^2(\beta) d\beta =1$. 
The notation for the roots has been kept the same as in Ref. 
\cite{IacX5}, while for the energies the notation $E_{s,n_w,L}$ 
will be used. The ground state band corresponds to $s=1$, $n_w=0$.
This model will be called the Z(4) model. 

The calculation of B(E2)s proceeds as described in Refs. \cite{Z4,Z5}.
The resulting level scheme is shown in Fig. 1(b). Experimental manifestations
of Z(4) seem to appear in the $^{128-132}$Xe region \cite{Z4}. 
One can easily see that the spectra of the ground state band and the 
$\beta_1$ band, as well as the related transitions, are very similar 
to the ones predicted by the E(5) model \cite{IacE5,E5}, while for the 
$\gamma_1$ bands the odd levels are very similar, while the even levels 
exhibit opposite staggering \cite{Z4}. A partial explanation of this 
behaviour is given in the next subsection. 

\subsection{Relation of the ground state band of Z(4) to E(4)} 

The ground state band of the Z(4) model is related to the second order Casimir 
operator of E(4), the Euclidean algebra in four dimensions. 
In order to see this, 
one can consider in general the Euclidean algebra in $n$ dimensions, E(n), 
which is the semidirect sum \cite{Wyb} of the algebra T$_n$ of translations 
in $n$ dimensions, generated by the momenta $P_j = -i \partial/ \partial x_j$, 
 and the SO(n) algebra 
of rotations in $n$ dimensions, generated by the angular momenta
\begin{equation}\label{eq:z53} 
 L_{jk} =-i \left(x_j{\partial \over \partial x_k} -x_k {\partial \over
\partial x_j} \right), 
\end{equation}
symbolically written as E(n) = T$_{\rm n}$ $\oplus_s$ SO(n) \cite{Barut}.  
The generators of E(n) satisfy the commutation relations 
\begin{equation}\label{eq:z54} 
 [P_i, P_j] =0, \qquad [P_i, L_{jk}] = i ( \delta_{ik} P_j - \delta_{ij} P_k),
\end{equation}
\begin{equation}\label{eq:z55} 
 [L_{ij}, L_{kl}]=i (\delta_{ik} L_{jl} +\delta_{jl} L_{ik} 
-\delta_{il} L_{jk} -\delta_{jk} L_{il}).
\end{equation}
From these commutation relations one can see that the square of the total 
momentum, $P^2$, is a second order Casimir operator of the algebra, while 
the eigenfunctions of this operator satisfy the equation 
\begin{equation}\label{eq:z56} 
 \left( -{1\over r^{n-1}} {\partial \over \partial r} r^{n-1} {\partial \over 
\partial r} + { \omega(\omega+n-2) \over r^2} \right) F(r) = k^2 F(r), 
\end{equation} 
in the left hand side of which the eigenvalues of the Casimir operator 
of SO(n), $\omega(\omega+n-2)$ appear \cite{Mosh1555}. 
Putting $ F(r) = r^{(2-n)/2} f(r)$ and $\nu= \omega+(n-2)/ 2$,
Eq. (\ref{eq:z56}) is brought into the form 
\begin{equation}\label{eq:z59} 
 \left( {\partial^2 \over \partial r^2} + {1\over r} {\partial \over \partial 
r} + k^2 - { \nu^2 \over r^2}\right) f(r)  =0, 
\end{equation}
the eigenfunctions of which are the Bessel functions $f(r) = J_\nu(kr) $.
The similarity between Eqs. (\ref{eq:z59}) and (\ref{eq:z47}) is clear. 

The ground state band of Z(4) is characterized by $n_w=0$, which means 
that $\alpha=L$. Then Eq. (\ref{eq:z48}) leads to $\nu=L/2+1$, while 
in the case of E(4) one has $\nu=\omega +1$. Then 
the two results coincide for $L=2\omega$, i.e. for even values of $L$. 
One can easily see that this coincidence occurs only in four dimensions. 

It should be emphasized, however, that neither the similarity of spectra 
and B(E2) values of Z(4) to these of the E(5) model, nor the coincidence
of the ground state band of Z(4) to the spectrum of the Casimir operator
of the Euclidean algebra E(4)
clarify the algebraic structure of the Z(4) model, the symmetry algebra of 
which has to be constructed explicitly, starting from the fact that $\gamma$ 
is fixed to $30^{\rm o}$, for which the Bohr Hamiltonian possesses 
``accidentally'' a symmetry axis (the body-fixed $\hat x'$-axis).

\section{The X(3) model} 

In the collective model of Bohr \cite{Bohr} the classical expression of the 
kinetic energy corresponding to $\beta$ and $\gamma$ vibrations of the nuclear
surface plus rotation of the nucleus has the form \cite{Bohr,ST}
\begin{equation}\label{eq:x1}
T = \frac{1}{2}\sum_{k=1}^{3} {\cal J}_k\, \omega^{\prime2}_k +
 \frac{B}{2}\,(\dot{\beta}^2+\beta^2 \dot{\gamma}^2), 
\end{equation}
where $\beta$ and $\gamma$ are the usual collective variables, $B$ is the 
mass parameter, 
$$ {\cal J}_k = 4B\beta^2 \sin^2\bigl(\gamma - 2\pi 
k/3\bigr)$$
are the three principal irrotational moments of inertia, 
and $\omega^\prime_k$ ($k=1$, 2, 3) are the components of the angular velocity
on the body-fixed $\hat k$-axes,
which can be expressed in terms of the time derivatives of the Euler angles 
$\dot{\phi}, \dot{\theta}, \dot{\psi}$ \cite{ST,Zare}  
\begin{equation}
\omega^\prime_1 = -\sin\theta \cos\psi\,\dot{\phi} + \sin\psi\,\dot{\theta},\, 
\omega^\prime_2 = \sin\theta \sin\psi\,\dot{\phi} + \cos\psi\,\dot{\theta},\,
\omega^\prime_3 = \cos\theta\,\dot{\phi} + \dot{\psi}.
\end{equation}
Assuming the nucleus to be $\gamma$-rigid (i.e. $\dot{\gamma}=0$), as in the 
Davydov and Chaban approach \cite{Chaban}, and considering in particular 
the axially symmetric prolate case of $\gamma=0$, we see that  
the third irrotational moment of inertia ${\cal J}_3$ vanishes, while the 
other two become equal (${\cal J}_1 = {\cal J}_2 = 3B\beta^2$), the kinetic 
energy of Eq. (\ref{eq:x1}) reaching the form \cite{ST,Davydov} 
\begin{equation}\label{eq:x4}
T = \frac{1}{2} 3B\beta^2 (\omega^{\prime2}_1 + \omega^{\prime2}_2) 
+ \frac{B}{2}\,\dot{\beta}^2
= \frac{B}{2}\Bigl[3\beta^2(\sin^2\theta\,\dot{\phi}^2 + \dot{\theta}^2) 
+ \dot{\beta}^2 \Bigr].
\end{equation}
It is clear that in this case the motion is characterized by three degrees 
of freedom. Introducing the generalized coordinates $q_1=\phi$, 
$q_2=\theta$, and $q_3=\beta$, the kinetic energy becomes 
a quadratic form of the time derivatives of the generalized coordinates
\cite{ST,EG}
$ T=B\sum_{i,j=1}^3 g_{ij}\; \dot{q}_i\dot{q}_j/2$, 
with the matrix $g_{ij}$ having a diagonal form
\begin{equation}\label{eq:x6} 
g_{ij} = \left(\begin{array}{ccc} 3\beta^2\sin^2\theta & 0 & 0 \\ 0 & 
3\beta^2 & 0 \\0 & 0 & 1 \end{array}\right). 
\end{equation}
(In the case of the full Bohr Hamiltonian \cite{Bohr} the square matrix 
$g_{ij}$ is 5-dimensional and non-diagonal \cite{ST,EG}.)

\begin{figure}[htb]
\includegraphics[height=100mm]{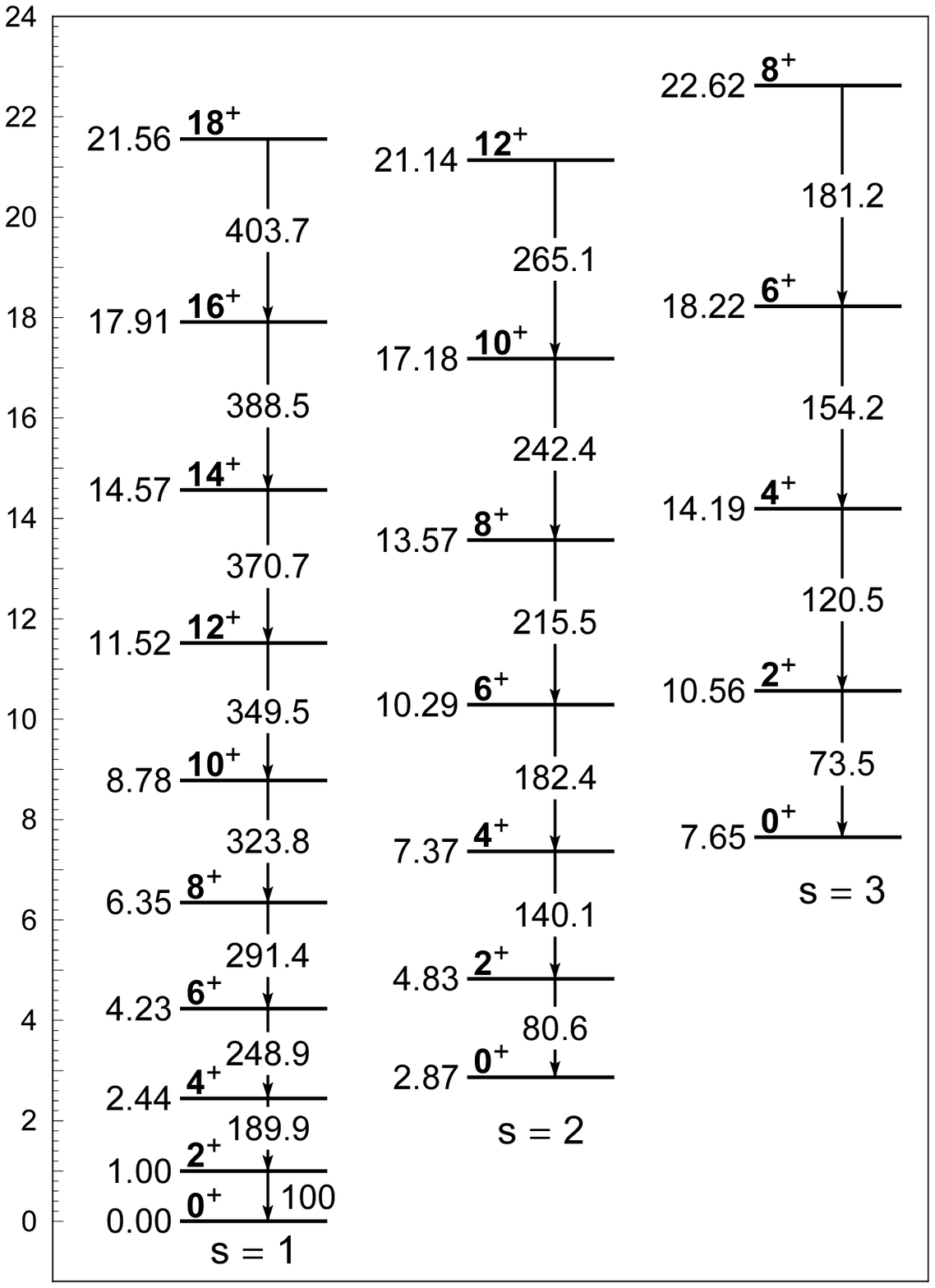} 
\caption{
Energy levels of the ground state ($s=1$), $\beta_1$ ($s=2$), and $\beta_2$ 
($s=3$) bands of X(3) \cite{X3}, normalized to the energy 
of the lowest excited state, $2_1^+$, together with 
intraband $B(E2)$ transition rates, normalized to the transition between 
the two lowest states, $B(E2; 2_1^+\to 0_1^+)$. Interband transitions 
are listed in Table 1.}
\label{fig2}
\end{figure}

Following the general procedure of quantization in curvilinear coordinates
one obtains the Hamiltonian operator \cite{ST,EG}
\begin{equation}
H = -\frac{\hbar^2}{2B}\,\Delta + U(\beta)
= -\frac{\hbar^2}{2B}\Biggl[\frac{1}{\beta^2}
\frac{\partial}{\partial\beta}\beta^2\frac{\partial}{\partial\beta} +
\frac{1}{3\beta^2} \Delta _\Omega\Biggr] + U(\beta), 
\end{equation}
where $\Delta_\Omega$ is the angular part of the Laplace operator
\begin{equation}
\Delta_\Omega = \frac{1}{\sin\theta}\frac{\partial}{\partial\theta}\sin\theta
\frac{\partial}{\partial\theta} + \frac{1}{\sin^2\theta}
\frac{\partial^2}{\partial\phi^2}. 
\end{equation}
The Schr\"odinger equation can be solved by the factorization
\begin{equation} \label{eq:x8}
\Psi(\beta,\theta,\phi) = F(\beta)\, Y_{LM}(\theta,\phi), 
\end{equation}
where $Y_{LM}(\theta,\phi)$ are the spherical harmonics. Then 
the angular part leads to the equation 
\begin{equation}\label{eq:9}
-\Delta_\Omega Y_{LM}(\theta,\phi) = L(L+1)Y_{LM}(\theta,\phi), 
\end{equation}
where $L$ is the angular momentum quantum number, 
while for the radial part $F(\beta)$ one obtains
\begin{equation}\label{eq:x9}
\Biggl[\frac{1}{\beta^2}
\frac{d}{d\beta}\beta^2\frac{d}{d\beta}
-\frac{L(L+1)}{3\beta^2} + \frac{2B}{\hbar^2}\Bigl(E-U(\beta)\Bigr)\Biggr] 
F(\beta) = 0. 
\end{equation}
As in the case of X(5) \cite{IacX5}, the potential in $\beta$ is taken to be 
an infinite square well [Eq. (\ref{eq:e7}]. 
In this case $F(\beta)$ is a solution of the equation
\begin{equation}\label{eq:x11}
\Biggl[\frac{d^2}{d\beta^2} + \frac{2}{\beta}\frac{d}{d\beta}
 + \Biggl(k^2-\frac{L(L+1)}{3\beta^2}\Biggr)\Biggr] F(\beta) = 0
\end{equation}
in the interval $0\leq\beta\leq\beta_W$, 
where reduced energies $\varepsilon=k^2=2B E/\hbar^2$ \cite{IacX5}
have been introduced, while it vanishes outside.

\begin{table}
\centering
\caption{Interband $B(E2; L_i\to L_f)$ transition rates for the X(3) model
\cite{X3}, normalized to the one between the two lowest states, 
$B(E2; 2_1^+\to 0_1^+)$.}
\medskip
\begin{tabular}{cr@{\qquad}cr@{\qquad}cr}
\hline\hline\noalign{\smallskip}
$L_i\to L_f$ & X(3) & $L_i\to L_f$ & X(3) & $L_i\to L_f$ & X(3) \\
\noalign{\smallskip}\hline\noalign{\smallskip}
$ 0_2 \to   2_1$ &164.0 &  & & & \\
$ 2_2 \to   4_1$ & 64.5 & $  2_2 \to   2_1$ &12.4 & $  2_2 \to   0_1$ & 0.54 \\
$ 4_2 \to   6_1$ & 42.2 & $  4_2 \to   4_1$ & 8.6 & $  4_2 \to   2_1$ & 0.43 \\
$ 6_2 \to   8_1$ & 31.1 & $  6_2 \to   6_1$ & 6.7 & $  6_2 \to   4_1$ & 0.51 \\
$ 8_2 \to  10_1$ & 24.4 & $  8_2 \to   8_1$ & 5.5 & $  8_2 \to   6_1$ & 0.56 \\
$10_2 \to  12_1$ & 19.9 & $ 10_2 \to  10_1$ & 4.7 & $ 10_2 \to   8_1$ & 0.59 \\
$ 0_3 \to   2_2$ &209.1 &  & & & \\
$ 2_3 \to   4_2$ & 92.0 & $  2_3 \to   2_2$ &16.2 & $  2_3 \to   0_2$ & 0.67 \\
$ 4_3 \to   6_2$ & 65.3 & $  4_3 \to   4_2$ &12.2 & $  4_3 \to   2_2$ & 0.47 \\
$ 6_3 \to   8_2$ & 50.9 & $  6_3 \to   6_2$ &10.1 & $  6_3 \to   4_2$ & 0.52 \\
$ 8_3 \to  10_2$ & 41.6 & $  8_3 \to   8_2$ & 8.6 & $  8_3 \to   6_2$ & 0.57 \\
$10_3 \to  12_2$ & 35.0 & $ 10_3 \to  10_2$ & 7.5 & $ 10_3 \to   8_2$ & 0.61 \\
\noalign{\smallskip}\hline\hline
\end{tabular}
\end{table}

Substituting $F(\beta)=\beta^{-1/2} f(\beta)$ one obtains the Bessel equation
\begin{equation}\label{eq:x12} 
\Biggl[\frac{d^2}{d\beta^2} +
\frac{1}{\beta}\frac{d}{d\beta}
 + \Biggl(k^2-\frac{\nu^2}{\beta^2}\Biggr)\Biggr]
 f(\beta) = 0,
\end{equation}
where 
$ \nu=\sqrt{\frac{L(L+1)}{3}+\frac{1}{4}}$, 
the boundary condition being $f(\beta_W)=0$. 
The solution of (\ref{eq:x11}), which is finite at $\beta=0$, is then 
\begin{equation}\label{eq:x14}
F(\beta)=F_{sL}(\beta) = \frac{1}{\sqrt{c}}\,\beta^{-1/2}
J_{\nu}(k_{s,\nu}\beta), 
\end{equation}
with $ k_{s,\nu}=x_{s,\nu}/\beta_W$ and $\varepsilon_{s,\nu}=k_{s,\nu}^2$,
where $x_{s,\nu}$ is the $s$-th zero of the Bessel function of the first kind 
$J_{\nu}(k_{s,\nu}\beta_W)$
and the normalization constant
$ c = \beta_W^2\,J^2_{\nu+1}(x_{s,\nu})/2$
is obtained from the condition 
$\int_{0}^{\beta_W}F_{sL}^2(\beta)\,\beta^2 d\beta = 1$.
The corresponding spectrum is then 
\begin{equation}\label{eq:x15} 
E_{s,L} = \frac{\hbar^2}{2B}\,k_{s,\nu}^2 =
\frac{\hbar^2}{2B\beta_W^2}\,x_{s,\nu}^2. 
\end{equation}
It should be noticed that in the X(5) case \cite{IacX5} 
the same Eq. (\ref{eq:x12}) occurs,
but with $\nu=\sqrt{ {L(L+1)\over 3} +{9\over 4} }$, while in the E(3) 
Euclidean algebra in 3 dimensions, which is the semidirect sum
of the T$_3$ algebra of translations in 3 dimensions and the SO(3) algebra of 
rotations in 3 dimensions \cite{Barut}, the eigenvalue equation of the square 
of the total momentum, which is a second-order Casimir operator of the 
algebra, also leads \cite{E5,Barut} to Eq. (\ref{eq:x12}), but with 
$\nu=L+{1\over 2}$.    

From the symmetry of the wave function of Eq. (\ref{eq:x8}) with respect 
to the plane which is orthogonal to the symmetry axis of the nucleus and goes 
through its center,
follows that the angular momentum $L$ can take only even nonnegative values.
Therefore no $\gamma$-bands appear in the model, as 
expected, since the $\gamma$ degree of freedom has been fixed 
to $\gamma=0$. 

B(E2) transition rates are calculated as described in Ref. \cite{X3}. 
The resulting level scheme is shown in Fig. 2 and Table 1. 
Experimental manifestations of X(3) seem to occur in $^{172}$Os and 
$^{186}$Pt \cite{X3}. An unexpected observation \cite{X3}
is that the $\beta_1$ bands 
of the N=90 isotones $^{150}$Nd, $^{152}$Sm, $^{154}$Gd, and $^{156}$Dy, 
agree very well with the X(3) predictions. These N=90 isotones are considered 
to be very good examples of X(5) \cite{CZX5,Kruecken,Tonev,Dewald,CaprioDy}, 
but the spacing within their $\beta_1$ bands 
is about half of that predicted by X(5). 

\section{Discussion}

It should be remarked that in all of the above mentioned cases the Bessel
eigenfunctions obtained are of the form $J_\nu(k \beta)$, with $\nu$ being 
of the form 
$ \nu=\sqrt{\Lambda + \left({n-2\over 2}\right)^2 }$,
where $n$ is the number of dimensions entering in the problem, while 
$\Lambda = L(L+1)/3$ in the cases of X(3) \cite{X3} and X(5) \cite{IacX5}, 
$\Lambda = [L(L+4)+3 n_w (2L-n_w)]/4$ in the cases of Z(4) \cite{Z4} and 
Z(5) \cite{Z5}, 
with $n_w =L-\alpha$ being the wobbling quantum number \cite{BM}, and 
$\Lambda = \tau(\tau+3)$ in the case of E(5), with $\tau$ being the seniority
quantum number characterizing the irreducible representations of the SO(5) 
subalgebra of E(5) \cite{IacE5}. In the corresponding ground state bands 
one has $n_w=0$ and $\tau=L/2$. 

It should also be mentioned that all the $\beta$-equations mentioned above 
are also soluble \cite{Elliott,Rowe} if the infinite square well potential is 
substituted by a Davidson 
potential \cite{Dav} of the form $u(\beta) =\beta^2 +\beta_0^4/\beta^2$, where 
$\beta_0$ is the minimum of the potential, the eigenfunctions being Laguerre 
polynomials instead of Bessel functions in this case. A variational procedure 
has been developed \cite{varPLB,varPRC}, in which the first derivative of 
various collective quantities
is maximized with respect to the parameter $\beta_0$, leading to the 
E(5), X(5), Z(5), and Z(4) results in the corresponding cases.
The solutions corresponding to the Davidson potentials lead to monoparametric 
curves \cite{PG} connecting various collective quantities, where agreement 
with experimental data is very good. 

Concerning future work, the clarification of the algebraic structure of the 
exactly soluble models X(3) and Z(4), as a prelude for the understanding 
of the algebraic structure of the approximate solutions X(5) and Z(5), 
is a challenging problem. The construction of analytical models including 
the octupole degree of freedom \cite{AQOA} and/or the dipole degree of freedom 
is also receiving attention. 

\section*{Acknowledgements}

One of the authors (IY) is thankful to the Turkish Atomic Energy Authority 
(TAEK) for support under project number 04K120100-4.

\end{document}